# On the photovoltaic effect asymmetry in ferroelectrics

S Semak[1,2], V Kapustianyk[1], Yu Eliyashevskyy[1], O Bovgyra[1], M Kovalenko[1], U Mostovoi[1], B Doudin[2] and B Kundys[2,*]

[1] Faculty of Physics, Ivan Franko National University of Lviv, Dragomanova 50, Lviv, UA79005, Ukraine
[2] Université de Strasbourg, CNRS, Institut de Physique et Chimie des Matériaux de Strasbourg, UMR 7504, 23 rue du Loess, F-67000 Strasbourg, France

**Abstract**

Despite symmetrical polarization, the magnitude of a light-induced voltage is known to be asymmetric with respect to poling sign in many photovoltaic (PV) ferroelectrics (FEs). This asymmetry remains unclear and is often attributed to extrinsic effects. We show here for the first time that such an asymmetry can be intrinsic, steaming from the superposition of asymmetries of internal FE bias and electro-piezo-strictive deformation. This hypothesis is confirmed by the observed decrease of PV asymmetry for smaller FE bias. Moreover, the both PV effect and remanent polarization are found to increase under vacuum-induced expansion and to decrease for gas-induced compression, with tens percents tunability. The change in cations positions under pressure is analysed through the first-principle density functional theory calculations. The reported properties provide key insight for FE-based solar elements optimization.

Keywords: strains, photovoltaicity, ferroelectrics, piezoelectric

## 1. Introduction

Harvesting solar energy with semiconductor-based photovoltaic (PV) cells is ultimately approaching its single bandgap fundamental limit [1, 2] pushing for increasingly complex multi-bandgap architectures. This issue can be overcome using ferroelectric (FE) PV materials where an ability of each single unit cell to act as a PV junction inspired research for decades. Historically the PV properties in non-centrosymmetric crystals were first reported by Chynoweth [3] for $BaTiO_3$. Two years later it was also observed in CdTe [4, 5] films and ZnS single crystals [6]. Because the photovoltages involved exceeded the bandgap energy of the material, the effect was fist termed 'anomalous photovoltaic effect' by Lempicki [7]. The reported photoconductivity in SbSI [8], electro-optical coupling [9] and ferroelectricity [10] allowed also to include this crystal [11] to the PV FE class. Due to its anisotropic nature and intrinsic origin the PV response was afterwards termed to be a 'bulk photovoltaic effect' by Glass *et al* [12]. Other PV FEs have been reported since then and theoretical models proposed [13–23]. A revival [24–28] in the field only occurred after the discovery of PV effects in multiferroic $BiFeO_3$ by Choi *et al* [29] triggering many other reports in new acentric compounds [26, 30–34]. The recent progress in PV efficiency in the $Bi_2FeCrO_6$ films [35] and other perovskite compounds [36–38], including 2D layers [39, 40], can make the FEs cells real competitors for conventional PVs in the near future. To make this happen, a better understanding of the physical mechanisms behind the PV response in FEs

* Author to whom any correspondence should be addressed.
*E-mail: kundysATipcms.fr

is essential. A compelling strategy to accomplish this is to find a model compound, preferably in a single crystal form to sort-out possible extrinsic effects occurring in thin films [41] and ceramics [42]. In this regard, we report here on the PV properties of the $Pb[(Mg_{1/3}Nb_{2/3})_{0.70}Ti_{0.30}]O_3$ (PMN-PT) FE crystal and focus our study on their correlation with its deformation properties. Our study is motivated by potential advantages development of FEs possessing above-bandgap photovoltages, stress sensitivity and electric field tunability.

## 2. Experimental detail

Some FE compounds of the PMN–PT family with the general formula of $Pb[Mg_{1/3}Nb_{2/3})_xTi_{1-x}]O_3$ were known to become PV after $WO_3$ doping [23] and recently the PV effect has been also reported for two compounds with stoichiometry at the vicinity of the morphotropic phase boundary region [33, 43, 44]. Crystals of this type are available commercially owing to their important electro-optical [45–47] and piezoelectric properties having a wide range of applications [48–50]. In this work a single crystal of $Pb[(Mg_{1/3}Nb_{2/3})_{0.70}Ti_{0.30}]O_3$ was ordered from Crystal GmbH (Germany) and its composition was independently verified by energy-dispersive x-ray spectroscopy (EDS). The EDS spectrum confirmed the absence of significant amounts of impurities and the exact content was found to be in a correspondence with the data of the supplier. Crystals of 0.3 mm in thickness, (001) orientation with edges along [010] and [100] were cut to form a plate 1.95 mm × 0.125 mm. The electrodes were made of silver paste and FE loops were measured along the longest sample dimension ([010]) using a homemade quasi-static FE tracer at 5.4 mHz. A light emitting diode M365FP1 (Thorlabs) at 356 nm wavelength and 9 nm bandwidth was used as a light source calibrated with a Thorlabs power meter (PM100USB). The light-intensity dependence was measured by increasing intensity with 2.97 mW cm$^{-2}$ steps taken at 330 ms intervals. The deformation was measured *in-situ* with the FE loop along the same poling direction [010] using a zigzag-type resistive strain gauge glued to the sample surface.

## 3. Results and discussions

### 3.1. PV asymmetry for positive and negative poling

To access the intrinsic PV properties and their poling dependence the hysteresis loop was first measured in darkness to check ferroelectricity and to stabilize specific FE remanent states (figure 1(a)). Because different FE states may exhibit different PV effects, this procedure must be well defined in order to ensure data reproducibility. If the FE loop is not measurable due to leakage or electric breakdown, at least the information on dielectric loss (or conductivity) along with pre-poling fields must accompany the PV characterization. In our case, the as–recorded FE loops demonstrated a well-defined hysteresis superposed to a minor leakage current. To obtain given remanent polarization values the sample was first depolarized by moving electric field to the negative coercive force and then back to zero. The maximum poling field was then increased stepwise for each individual minor FE loop until reaching over-coercive values. Since the build-in electric field governs the photo-carriers separation, different FE states formed by sub-coercive poling should have different PV responses. For demonstration purposes the light intensity dependence of the open circuit photovoltage ($V_{oc}$) was therefore measured at several different remanent FE states for sub-coercive and overcoercive regions (figure 1(b)).

First of all, a small PV trend change for lower intensities (<25 mW cm$^{-2}$) related to pyroelectric effect [3, 15] and depolarization [45, 51] was observed. As light intensity increased a clear poling dependence of PV effect becomes more evident. After positive poling the PV effect (remanent states 1–8) was generally two times larger comparing to negative poling. For instance, the absolute ratio of the PV response taken between the two remanent points 1 and 16 at 250 mW cm$^{-2}$ light intensity reached 1.94 value. Due to the non-centrosymmetric structure of an ideal FE crystal the atomic motions induced by saturating electric-field relate to 180° dipole reversal. The remanent polarization states $+P_r$ (state 1) and $-P_r$ (state 16) obtained after positive and negative poling should have therefore equal absolute values. In striking contrast, the absolute value of the PV response depends significantly on the poling sign for the both subcoercive and overcoercive regions (figure 1(b)). Such an asymmetry with respect to the poling sign has been reported in many FEs [26, 52–55] and was generally attributed to asymmetric Schottky barrier height or inhomogeneous distribution of charged defects. Here, however, we deal with a single crystal with symmetrical electrodes and perpendicular geometry of the experiment designed to exclude a need of a transparent electrode. Therefore, the observed poling sign dependence of the PV effect must have an intrinsic origin.

### 3.2. Electrically-induced remanent strain for positive and negative poling

In order to explain the PV asymmetry one has to recall that the origin of the PV effect in FEs is in their non-centrosymmetric structure [21] that allows the generated photocarriers to not immediately recombine. On the other hand, the electric driven change in electric polarization (i.e. primary order parameter) can be accompanied by a lattice deformation (secondary order parameter). Such deformation in FEs can exhibit both linear (piezoelectric) and quadratic (electrostrictive) contributions (figure 2(d) (inset)). While the first one is specific to 20 symmetry crystal classes, the last one can exist in any insulating solid, including a polar one. The combination of the linear and quadratic strain effects can lead to asymmetric branches in strain versus electric field. This asymmetry can be further increased by the asymmetry in coercive electric fields after positive and negative poling known as internal bias field (figure 2(c) (inset)). As a result, the measured deformation along the poled direction exhibits remanent tension after positive poling and remanent compression after negative poling with respect to fully depolarized state (figure 2(b)). The resulting piezoelectric loop is therefore not symmetric (figure 2(d)) as often reported for many piezoelectric samples [56–60]. From the comparison between deformation measurements and PV responses it is evident that a tension along the polarization direction (state 1, obtained after positive poling) is more favourable for PV effect, in contrast to compression (state 16). It has to be noted that the composition of our sample is known to be a mixture of orthorhombic and rhombohedral structures [61, 62] where the rhombohedral phase can dominate [63–65].

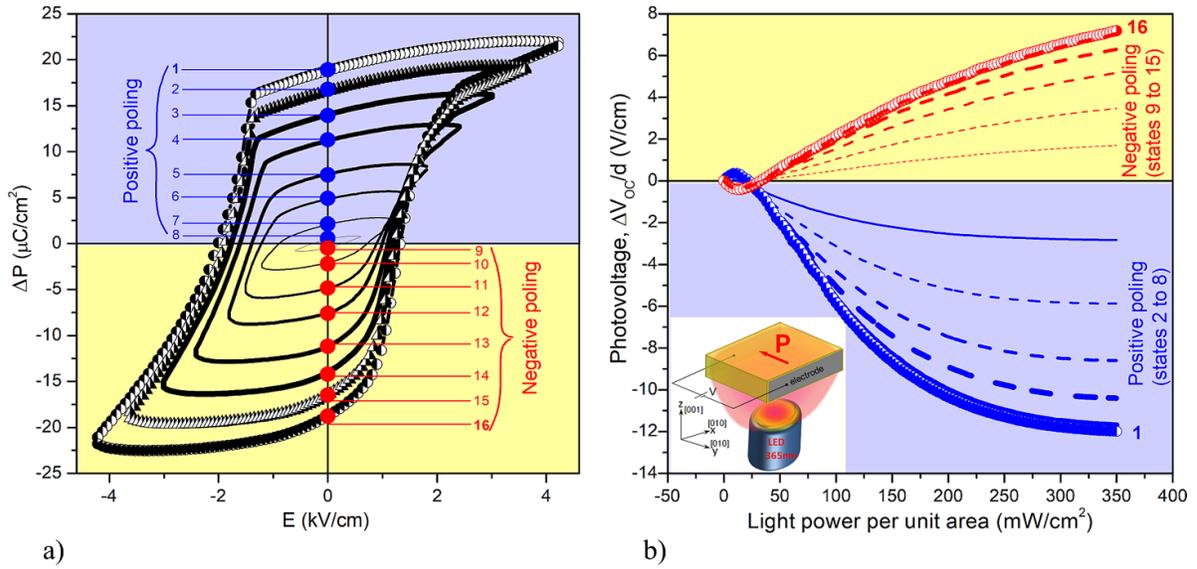

**Figure 1.** Asymmetricity of PV effect with respect to poling sign. (a) Different remanent states of FE loop for the Pb[(Mg$_{1/3}$Nb$_{2/3}$)$_{0.7}$Ti$_{0.30}$]O$_3$ crystal in darkness. (b) Intensity dependence of the open circuit photovoltage (V$_{oc}$) divided by the distance between electrodes for the same remanent FE states. States from 2 to 15 schematically demonstrate the poling dependent PV effect. The inset illustrates the experimental scheme.

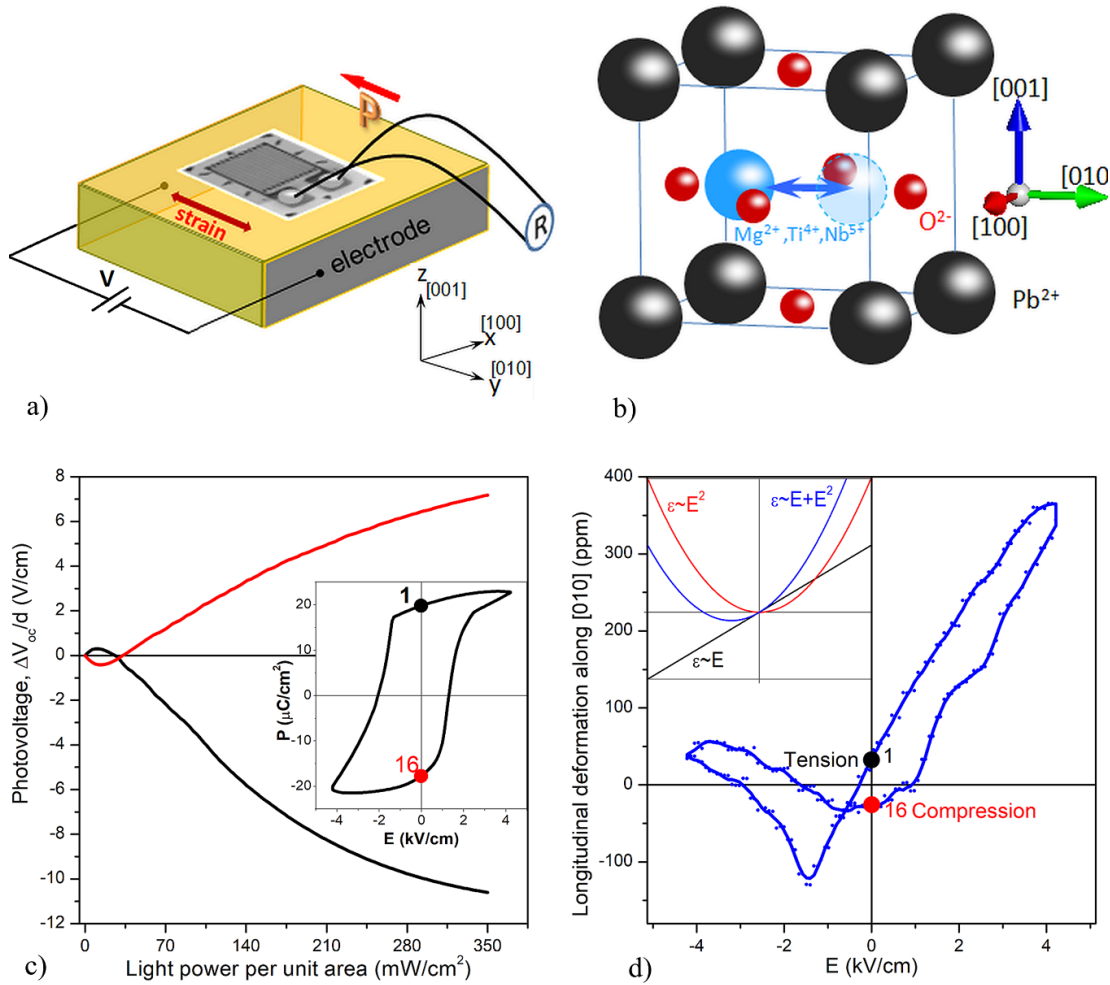

**Figure 2.** (a) Schematic illustration of the longitudinal strain measurements in darkness. (b) Rhombohedral unit cell of the Pb[(Mg$_{1/3}$Nb$_{2/3}$)$_{0.70}$Ti$_{0.30}$]O$_3$ crystal. (c) Intensity dependence of the photovoltage measured for opposite remanent states of the FE loop (inset). (d) The deformation loop with two different remanent deformation states obtained after positive and negative poling. The inset to figure (d) explains the deformation asymmetry due to quadratic and linear contributions.

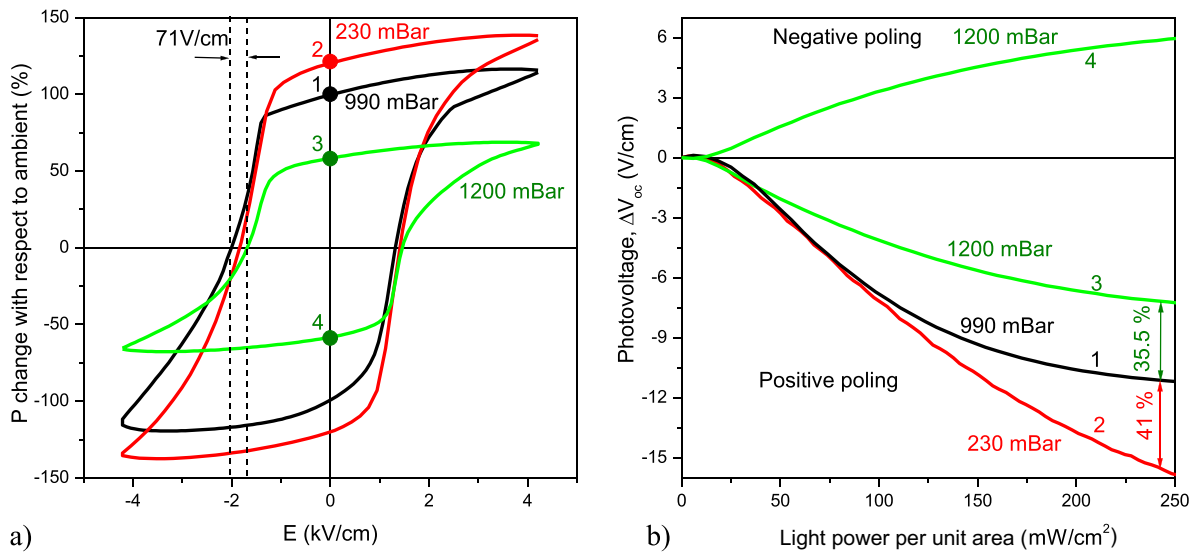

**Figure 3.** (a) FE loop under different uniform pressure conditions. (b) Photovoltage versus light intensity measured for atmospheric pressure, compression and vacuum.

Because our crystal was previously polarized along the [010] direction (figure 2(b)), tension along this direction should increase polarizability, while compression should have an opposite effect. This conclusion applies to both the mentioned internal field and the externally applied mechanical stresses. Indeed, the influence of a compressive stress on polarization in PMN-30PT single crystal was observed in [66]. Since crystalline acentricity is at the origin of PV effect [21] in FEs, the photoresponse can decrease for negative stress that reduces the acentricity and increase for the opposite.

### 3.3. Mechanically-induced strain by vacuum and gas pressures

To verify the strain control hypothesis over PV effect asymmetry due to poling sign the mechanical way to apply deformation to the sample was tested. As the uniaxial tension is technically difficult to apply to mm-size samples, our crystal was placed into the optical cryostat and exposed to vacuum to obtain lattice expansion and gas pressure to obtain lattice compression. First of all, the tension results in a remarkable increase of the remanent polarization by ∼20% and also increases the PV effect by ∼41% at the 250 mW cm$^{-2}$ of light intensity, confirming our hypothesis (figure 3). To verify the impact of compression we have increased pressure up to 1200 mBar by injecting $N_2$ gas. The action of compression indeed results in the decrease of polarization (by ∼40%) and PV effect (by ∼35.5%) for the same light intensity. Interestingly, we also observe a modification of the FE internal bias field when changing the pressure. The most pronounced reduction in the FE bias field of 71 V m$^{-1}$ is observed under 1200 mBar pressure (green FE loop in figure 3(a)). This experimental observation allows us to independently verify the PV asymmetry dependence on the internal bias. Indeed, the smaller FE bias field by 71 V m$^{-1}$ for 1200 mBar FE loop leads to a reduction of the PV asymmetry more than doubled when compared to ambient pressure.

The absolute ratio of PV response taken at the FE remanence (for the points 3 and 4 at 250 mW cm$^{-2}$ in figure 3(b)) is now only 1.21, i.e. 60% more symmetric than for ambient pressure (the larger FE bias field) (figure 1(b)). It is therefore confirmed that the asymmetry in PV effect arises from the strain correlated FE bias field i.e. asymmetry in coercive voltages, while the higher PV effect is correlated with higher polarizability.

### 3.4. Density functional theory calculations

The deformation-dependent polarization of the local atomic environment can be modelled using density functional theory (DFT) where both hydrostatic and uniaxial deformation can be considered with a stress of a different sign. For sake of simplicity the first-principles calculations were performed for a 2 × 2 × 2 supercell with 40 atoms to study the local structures of the $Pb[(Mg_{1/3}Nb_{2/3})_{0.75}Ti_{0.25}]O_3$ (0.75PMN–0.25PT), which has been successfully applied in previous studies [67, 68] and was established to be large enough to describe the local structure properties. The periodic DFT calculations were carried out using the plane-wave pseudopotential method as implemented in the CASTEP code

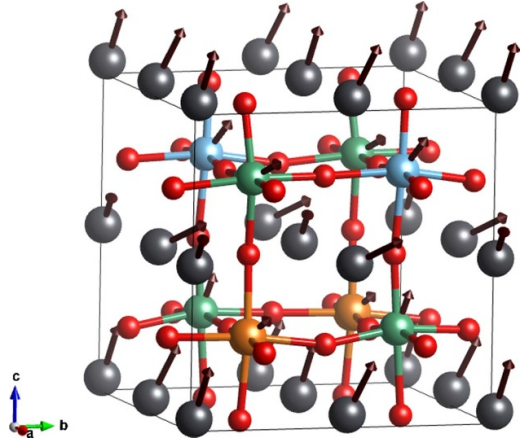

**Figure 4.** Relaxed structure of 40-atoms 2 × 2 × 2 supercell of 0.75PMN–0.25PT obtained by DFT calculations. Arrows (with a scale factor of 3 for better visualization) show the displacements from high-symmetry perovskite positions. Gray, green, orange, blue, and red spheres represent Pb, Nb, Mg, Ti, and O atoms, respectively.

**Table 1.** Local structure parameters of 0.75PMN–0.25PT model system at equilibrium state and under an external stress (hydrostatic pressure and uniaxial strain), $V$ is the volume of the 2 × 2 × 2 40-atom supercell (in Å$^3$), $D_{Pb}$, $D_{Mg}$, $D_{Nb}$, and $D_{Ti}$ are the average displacements of Pb$^{2+}$, Mg$^{2+}$, Nb$^{5+}$, and Ti$^{4+}$ cations respectively (in Å).

| Stress | $V$, Å$^3$ | $D_{Pb}$ | $D_{Mg}$ | $D_{Nb}$ | $D_{Ti}$ |
|---|---|---|---|---|---|
| 0 GPa | 510.431 | 0.395 | 0.076 | 0.177 | 0.220 |
| Hydrostatic pressure 0.05 GPa | 510.014 | 0.390 | 0.076 | 0.175 | 0.218 |
| Hydrostatic pressure 0.10 GPa | 509.606 | 0.385 | 0.076 | 0.174 | 0.216 |
| Negative hydrostatic pressure −0.10 GPa | 510.694 | 0.398 | 0.077 | 0.179 | 0.222 |
| Uniaxial compression [100] direction, 0.10 GPa | 510.128 | 0.391 | 0.076 | 0.175 | 0.218 |
| Uniaxial compression [010] direction, 0.10 GPa | 510.290 | 0.395 | 0.076 | 0.177 | 0.220 |
| Uniaxial compression [001] direction, 0.10 GPa | 510.030 | 0.388 | 0.077 | 0.175 | 0.218 |
| Uniaxial extension [100] direction, −0.10 GPa | 510.344 | 0.390 | 0.077 | 0.176 | 0.219 |
| Uniaxial extension [010] direction, −0.10 GPa | 510.336 | 0.390 | 0.076 | 0.175 | 0.219 |
| Uniaxial extension [001] direction, −0.10 GPa | 510.617 | 0.397 | 0.076 | 0.178 | 0.221 |

[69]. Exchange and correlation were approximated using the GGA PBE functional. The plane-wave cut-off energy was 800 eV. The Brillouin zone sampling of electronic states was performed on a 4 × 4 × 4 Monkhorst-Pack grid. The equilibrium crystal structures were obtained via geometry optimization in the Broyden-Fletcher-Goldfarb-Shanno minimization scheme after which the residual forces were converged to zero within 0.005 eV Å$^{-1}$. The optimised lattice parameters for 0.75PMN–0.25PT supercell (figure 4) are as follows: $a = 8.043$ Å, $b = 7.890$ Å, $c = 8.043$ Å, $\alpha = 90.015°$, $\beta = 89.632°$, $\gamma = 90.022°$ and volume $V = 510.431$ Å$^3$(63.8 Å$^3$/f.u.).

The values of Pb and B-cation (Mg, Nb, Ti) displacements inside their respective oxygen cages are the determining factor for the macroscopic properties such as polarization in PMN-PT FE perovskites. The Pb$^{2+}$ 6$s$ lone-pair electrons show a large Pb displacement and therefore a significant Pb contribution to the overall polarization [67].

Results of DFT calculations for the average displacement magnitudes of Pb$^{2+}$, Mg$^{2+}$, Nb$^{5+}$, and Ti$^{4+}$ cations from the negative charge centre of its surrounding oxygen atoms in 0.75PMN–0.25PT model system are presented in table 1 and indicated by arrows in figure 4. The calculated unit cell parameters and local structure parameters show a good agreement with previous theoretical and experimental results [67, 68, 70].

Analysis of the local structure at equilibrium state shows that the Pb ions undergo the largest displacement of 0.395 Å from their initial positions, moving towards the Mg–Nb face [(001) plane] and avoiding the Ti–Nb face [(001) plane], due to the stronger repulsive force between Pb–Nb and Pb–Ti than the one between Pb–Mg. The Nb and Ti ions move along the same direction as Pb, and Mg ions are only slightly displaced. It is seen that both Nb and Ti atoms move off-centre by 0.18–0.22 Å and create shorter Nb–O and Ti–O bonds, making decisive contribution to the macroscopic polarization [71].

In the PMN-PT system under hydrostatic compression Pb, Nb and Ti displacement magnitudes are reduced and the structure becomes more centrosymmetric with an expected smaller PV effect. When we apply a negative hydrostatic pressure to our model system, we obtain the expected increase of the unit cell volume and enhanced amplitude of the $Pb^{2+}$ displacements along with $Nb^{5+}$ and $Ti^{4+}$ cations. The major difference between the calculations results for the hydrostatic pressure and uniaxial compression, as well as for the negative hydrostatic pressure and uniaxial extension, is a clear anisotropy of cation displacement magnitudes, which should lead to anisotropy of the stress-induced PV behaviour of the PMN-PT crystal. Examination of the data in table 1 for uniaxial compression and tension shows that all cation displacement magnitudes predicted by DFT remain unchanged, when the system is under uniaxial compression along [010] direction with the lattice parameters and the unit cell volume changing accordingly. Hence, in this case the PV response should remain unchanged. The largest decrease of the average magnitude of the Pb-cation displacements is obtained under compression along the [001] direction, which should result in decreasing of PV effect. Indeed, the significant reduction of remanent polarization under compressive uniaxial stress along [001] in PMN-PT crystal was observed [72]. Stretching along [100] and [010] axes gives a similar result of reducing the displacement of Pb ions, while stretching along [001] direction leads to a larger magnitude of Pb, Nb, and Ti distortions and should enhance the PV effects.

## 4. Conclusions

In conclusion, a direct correlation between the magnitude of PV effect and a sign of electrically induced stress along the measured direction has been observed for a PV and FE crystal. This correlation is proposed to resolve a long standing question on PV magnitude asymmetry with respect to the poling sign in PV FEs. The internal FE bias field resulting in remanent tension or compression after poling strongly impacts the PV response. Tension results in an increase of PV effect along the polarization direction, while compression does the opposite. Because intrinsic crystalline acentricity is affected by strain the both magnitude of FE polarization and PV effect can be tuned by modifying the inert gas pressure environment of the sample. While larger FE bias field leads to larger PV asymmetry, the larger polarizability is favourable for larger PV effect. One should also emphasize the remarkable sensitivity of the polarization and the resulting PV amplitude to external pressure, with a relative variation of the same order of magnitude as the relative pressure change around ambient. The important role of strain for the PV effect in FEs is now progressively recognized in solid state physics where the both enhancement [73] and creation [74, 75] of PV effects were reported. Considering possible existence of multiple sub-coercive electric states, the single domain state with the accompanied remanent tension along the polarization direction can be favourable for larger PV response in a poled direction. Taking into account that internal bias field giving rise to opposite remanent stresses can be also formed by growth conditions, thermal/electrical [76, 77] history or doping [78], our results on the well-controlled model material can explain differences or lack of reproducibility in the literature. Regardless the way of stress introduction: either material fabrication [79, 80] or post-processing techniques, the strain engineering can be considered as a key enabling strategy for tuning and optimizing PV properties in FEs.

## Acknowledgments

This work was supported by the Campus France Project (Dnipro No. 46789RH) and by the Ministry of Education and Science of Ukraine. We thank U Burkhardt and T Ferté for EDS analysis. S S acknowledges PAUSE program from the French ministry of higher education and research. The technical help of F Chevrier and discussions with H Majjad and M Lenertz are also acknowledged.

## References

[1] Huang J, Yuan Y, Shao Y and Yan Y 2017 Understanding the physical properties of hybrid perovskites for photovoltaic applications *Nat. Rev. Mater.* **2** 17042
[2] Shockley W and Queisser H J 1961 Detailed balance limit of efficiency of p-n junction solar cells *J. Appl. Phys.* **32** 510–9
[3] Chynoweth A G 1956 Surface space-charge layers in barium titanate *Phys. Rev.* **102** 705–14
[4] Pensak L 1958 High-voltage photovoltaic effect *Phys. Rev.* **109** 601
[5] Goldstein B 1958 Properties of photovoltaic films of CdTe *Phys. Rev.* **109** 601–3
[6] Ellis S G, Herman F, Loebner E E, Merz W J, Struck C W and White J G 1958 Photovoltages larger than the band gap in zinc sulfide crystals *Phys. Rev.* **109** 1860

[7] Lempicki A 1959 Anomalous photovoltaic effect in ZnS single crystals *Phys. Rev.* **113** 1204–9

[8] Nitsche R and Merz W J 1960 Photoconduction in ternary V-VI-VII compounds *J. Phys. Chem. Solids* **13** 154–5

[9] Kern R 1962 An electro-optical and electromechanical effect in SbSI *J. Phys. Chem. Solids* **23** 249–53

[10] Fatuzzo E, Harbeke G, Merz W J, Nitsche R, Roetschi H and Ruppel W 1962 Ferroelectricity in SbSI *Phys. Rev.* **127** 2036–7

[11] Fridkin V M, Grekov A A, Rodin A I, Savchenko E A and Volk T R 1973 Photoconductivity in ferroelectrics *Ferroelectrics* **6** 71–82

[12] Glass A M, von der Linde D and Negran T J 1974 High-voltage bulk photovoltaic effect and the photorefractive process in $LiNbO_3$ *Appl. Phys. Lett.* **25** 233–5

[13] Fridkin V M and Popov B N 1978 Anomalous photovoltaic effect in ferroelectrics *Sov. Phys., Usp.* **21** 981–91

[14] Kraut W and von Baltz R 1979 Anomalous bulk photovoltaic effect in ferroelectrics: a quadratic response theory *Phys. Rev. B* **19** 1548–54

[15] Chen F S 1969 Optically induced change of refractive indices in $LiNbO_3$ and $LiTaO_3$ *J. Appl. Phys.* **40** 3389–96

[16] Belinicher V I and Sturman B I 1980 The photogalvanic effect in media lacking a center of symmetry *Sov. Phys., Usp.* **23** 199

[17] Fridkin V M 2001 Bulk photovoltaic effect in noncentrosymmetric crystals *Crystallogr. Rep.* **46** 654–8

[18] Sturman B and Fridkin V M 1992 *Photovoltaic and Photo-refractive Effects in Noncentrosymmetric Materials* 1st edn (London: Routledge) p 264

[19] Fridkin V M, Popov B N, Kuznetsov V A and Barsukova M L 2011 The photoconductivity and photovoltaic effect in $PbTiO_3$ *Ferroelectrics* **422** 70–76

[20] Bune A V, Fridkin V M, Verkhovskaya K A and Taylor G 1990 Photoelectric properties of the ferroelectric polymer poly(vinylidene fluoride) *Polym. J.* **22** 7–14

[21] von Baltz R and Kraut W 1981 Theory of the bulk photovoltaic effect in pure crystals *Phys. Rev. B* **23** 5590–6

[22] Ichiki M, Furue H, Kobayashi T, Maeda R, Morikawa Y, Nakada T and Nonaka K 2005 Photovoltaic properties of $(Pb,La)(Zr,Ti)O_3$ films with different crystallographic orientations *Appl. Phys. Lett.* **87** 222903

[23] Tu C-S, Wang F-T, Chien R R, Schmidt V H, Hung C-M and Tseng C-T 2006 Dielectric and photovoltaic phenomena in tungsten-doped $Pb(Mg_{1/3}Nb_{2/3})_{1-x}Ti_xO_3$ crystal *Appl. Phys. Lett.* **88** 032902

[24] Butler K T, Frost J M and Walsh A 2015 Ferroelectric materials for solar energy conversion: photoferroics revisited *Energy Environ. Sci.* **8** 838–48

[25] Paillard C, Bai X, Infante I C, Guennou M, Geneste G, Alexe M, Kreisel J and Dkhil B 2016 Photovoltaics with ferroelectrics: current status and beyond *Adv. Mater.* **28** 5153–68

[26] Grinberg I *et al* 2013 Perovskite oxides for visible-light-absorbing ferroelectric and photovoltaic materials *Nature* **503** 509

[27] Chen G, Chen J, Pei W, Lu Y, Zhang Q, Zhang Q and He Y 2019 Bismuth ferrite materials for solar cells: current status and prospects *Mater. Res. Bull.* **110** 39–49

[28] Han X, Ji Y and Yang Y 2022 Ferroelectric photovoltaic materials and devices *Adv. Funct. Mater.* **32** 2109625

[29] Choi T, Lee S, Choi Y J, Kiryukhin V and Cheong S-W 2009 Switchable ferroelectric diode and photovoltaic effect in $BiFeO_3$ *Science* **324** 63–66

[30] Kumari S, Ortega N, Kumar A, Scott J F and Katiyar R S 2014 Ferroelectric and photovoltaic properties of transition metal doped $Pb(Zr_{0.14}Ti_{0.56}Ni_{0.30})O_{3-\delta}$ thin films *AIP Adv.* **4** 037101

[31] Chakrabartty J P, Nechache R, Harnagea C and Rosei F 2014 Photovoltaic effect in multiphase Bi–Mn–O thin films *Opt. Express* **22** A80–9

[32] Pérez-Tomás A, Lira-Cantú M and Catalan G 2016 Above-bandgap photovoltages in antiferroelectrics *Adv. Mater.* **28** 9644–7

[33] Makhort A S, Chevrier F, Kundys D, Doudin B and Kundys B 2018 Photovoltaic effect and photopolarization in $Pb[(Mg_{1/3}Nb_{2/3})_{0.68}Ti_{0.32}]O_3$ crystal *Phys. Rev. Mater.* **2** 012401

[34] Wu L, Podpirka A, Spanier J E and Davies P K 2019 Ferroelectric, optical, and photovoltaic properties of morphotropic phase boundary compositions in the $PbTiO_3$–$BiFeO_3$–$Bi(Ni_{1/2}Ti_{1/2})O_3$ system *Chem. Mater.* **31** 4184–94

[35] Nechache R, Harnagea C, Li S, Cardenas L, Huang W, Chakrabartty J and Rosei F 2015 Bandgap tuning of multiferroic oxide solar cells *Nat. Photonics* **9** 61–67

[36] Yang M, Zhou Y, Zeng Y, Jiang C-S, Padture N P and Zhu K 2015 Square-centimeter solution-processed planar $CH_3NH_3PbI_3$ perovskite solar cells with efficiency exceeding 15% *Adv. Mater.* **27** 6363–70

[37] Zheng X *et al* 2020 Managing grains and interfaces via ligand anchoring enables 22.3%-efficiency inverted perovskite solar cells *Nat. Energy* **5** 131–40

[38] Li L *et al* 2022 Flexible all-perovskite tandem solar cells approaching 25% efficiency with molecule-bridged hole-selective contact *Nat. Energy* **7** 1–10

[39] Yang D *et al* 2022 Spontaneous-polarization-induced photovoltaic effect in rhombohedrally stacked $MoS_2$ *Nat. Photon.* **16** 469–74

[40] Jiang J, Chen Z, Hu Y, Xiang Y, Zhang L, Wang Y, Wang G-C and Shi J 2021 Flexo-photovoltaic effect in $MoS_2$ *Nat. Nanotechnol.* **16** 894–901

[41] Calzada M L, Jiménez R, González A, García-López J, Leinen D and Rodríguez-Castellón E 2005 Interfacial phases and electrical characteristics of ferroelectric strontium bismuth tantalate films on $Pt/TiO_2$ and Ti/Pt/Ti heterostructure electrodes *Chem. Mater.* **17** 1441–9

[42] Takagi K, Kikuchi S, Li J-F, Okamura H, Watanabe R and Kawasaki A 2004 Ferroelectric and photostrictive properties of fine-grained PLZT ceramics derived from mechanical alloying *J. Am. Ceram. Soc.* **87** 1477–82

[43] Makhort A S, Schmerber G and Kundys B 2019 Larger photovoltaic effect and hysteretic photocarrier dynamics in $Pb[(Mg_{1/3}Nb_{2/3})_{0.70}Ti_{0.30}]O_3$ crystal *Mater. Res. Express* **6** 066313

[44] Liew W H, Chen Y, Alexe M and Yao K 2022 Fast photostriction in ferroelectrics *Small* **18** 2106275

[45] Makhort A, Gumeniuk R, Dayen J-F, Dunne P, Burkhardt U, Viret M, Doudin B and Kundys B 2022 Photovoltaic-ferroelectric materials for the realization of all-optical devices *Adv. Opt. Mater.* **10** 2102353

[46] Liu X *et al* 2022 Ferroelectric crystals with giant electro-optic property enabling ultracompact Q-switches *Science* **376** 371–7

[47] Qiu C *et al* 2020 Transparent ferroelectric crystals with ultrahigh piezoelectricity *Nature* **577** 350–4

[48] Mansour R, Omoniyi O A, Reid A, Brindley W, Stewart B G and Windmill J F C 2022 Synergy of PMN-PT with piezoelectric polymer using sugar casting method for sensing applications *2022 IEEE Int. Conf. on Flexible and Printable Sensors and Systems (FLEPS)* pp 1–4

[49] Zhang T *et al* 2022 Piezoelectric ultrasound energy–harvesting device for deep brain stimulation and analgesia applications *Sci. Adv.* **8** eabk0159

[50] Yamashita Y, Karaki T, Lee H-Y, Wan H, Kim H-P and Jiang X 2022 A review of lead perovskite piezoelectric single crystals and their medical transducers application

  *IEEE Trans. Ultrason. Ferroelectr. Freq. Control*
  **69** 3048–56
- [51] Li T *et al* 2018 Optical control of polarization in ferroelectric heterostructures *Nat. Commun.* **9** 3344
- [52] Park J, Won S S, Ahn C W and Kim I W 2013 Ferroelectric photocurrent effect in polycrystalline lead-free $(K_{0.5}Na_{0.5})(Mn_{0.005}Nb_{0.995})O_3$ thin film *J. Am. Ceram. Soc.* **96** 146–50
- [53] Pintilie L, Dragoi C and Pintilie I 2011 Interface controlled photovoltaic effect in epitaxial Pb(Zr,Ti)O$_3$ films with tetragonal structure *J. Appl. Phys.* **110** 044105
- [54] Lan Y *et al* 2022 Achieving ultrahigh photocurrent density of Mg/Mn-modified KNbO$_3$ ferroelectric semiconductors by bandgap engineering and polarization maintenance *Chem. Mater.* **34** 4274–85
- [55] Rani K, Matzen S, Gable S, Maroutian T, Agnus G and Lecoeur P 2021 Quantitative investigation of polarization-dependent photocurrent in ferroelectric thin films *J. Phys. Condens. Matter* **34** 104003
- [56] Cornelius T W, Mocuta C, Escoubas S, Lima L R M, Araújo E B, Kholkin A L and Thomas O 2020 Piezoelectric properties of $Pb_{1-x}La_x(Zr_{0.52}Ti_{0.48})_{1-x/4}O_3$ thin films studied by *in situ* x-ray diffraction *Materials* **13** 3338
- [57] Rodríguez-Aranda M C, Calderón-Piñar F, Espinoza-Beltrán F J, Flores-Ruiz F J, León-Sarabia E, Mayén-Mondragón R and Yáñez-Limón J M 2014 Ferroelectric hysteresis and improved fatigue of PZT (53/47) films fabricated by a simplified sol–gel acetic-acid route *J. Mater. Sci.: Mater. Electron.* **25** 4806–13
- [58] Damjanovic D 1998 Ferroelectric, dielectric and piezoelectric properties of ferroelectric thin films and ceramics *Rep. Prog. Phys.* **61** 1267–324
- [59] Yoon I T, Lee J, Tran N C and Yang W 2020 Polarity control of ZnO films grown on ferroelectric (0001) LiNbO$_3$ substrates without buffer layers by pulsed-laser deposition *Nanomaterials* **10** 380
- [60] Song J M, Luo L H, Dai X H, Song A Y, Zhou Y, Li Z N and Liu B T 2018 Switching properties of epitaxial $La_{0.5}Sr_{0.5}CoO_3/Na_{0.5}Bi_{0.5}TiO_3/La_{0.5}Sr_{0.5}CoO_3$ ferroelectric capacitor *RSC Adv.* **8** 4372–6
- [61] Guo Y, Luo H, Ling D, Xu H, He T and Yin Z 2003 The phase transition sequence and the location of the morphotropic phase boundary region in (1−x)[Pb(Mg$_{1/3}$Nb$_{2/3}$)O$_3$]–xPbTiO$_3$ single crystal *J. Phys. Condens. Matter* **15** L77
- [62] Guo Y, Luo H, Chen K, Xu H, Zhang X and Yin Z 2002 Effect of composition and poling field on the properties and ferroelectric phase-stability of Pb(Mg$_{1/3}$Nb$_{2/3}$)O$_3$–PbTiO$_3$ crystals *J. Appl. Phys.* **92** 6134–8
- [63] Long X and Ye Z-G 2007 Top-seeded solution growth and characterization of rhombohedral PMN–30PT piezoelectric single crystals *Acta Mater.* **55** 6507–12
- [64] Cao H, Bai F, Wang N, Li J, Viehland D, Xu G and Shirane G 2005 Intermediate ferroelectric orthorhombic and monoclinic $M_B$ phases in [110] electric-field-cooled Pb(Mg$_{1/3}$Nb$_{2/3}$)O$_3$–30% PbTiO$_3$ crystals *Phys. Rev.* B **72** 064104
- [65] Bai F, Wang N, Li J, Viehland D, Gehring P M, Xu G and Shirane G 2004 X-ray and neutron diffraction investigations of the structural phase transformation sequence under electric field in 0.7Pb(Mg$_{1/3}$Nb$_{2/3}$)-0.3PbTiO$_3$ crystal *J. Appl. Phys.* **96** 1620–7
- [66] Zhang H 2014 Effect of the compressive stress on both polarization rotation and phase transitions in PMN-30%PT single crystal *AIP Adv.* **4** 057118
- [67] Tan H, Takenaka H, Xu C, Duan W, Grinberg I and Rappe A M 2018 First-principles studies of the local structure and relaxor behavior of Pb(Mg$_{1/3}$Nb$_{2/3}$)O$_3$–PbTiO$_3$-derived ferroelectric perovskite solid solutions *Phys. Rev.* B **97** 174101
- [68] Li C, Xu B, Lin D, Zhang S, Bellaiche L, Shrout T R and Li F 2020 Atomic-scale origin of ultrahigh piezoelectricity in samarium-doped PMN-PT ceramics *Phys. Rev.* B **101** 140102
- [69] Clark S J, Segall M D, Pickard C J, Hasnip P J, Probert M I J, Refson K and Payne M C 2005 First principles methods using CASTEP *Z. Kristallogr. Cryst. Mater.* **220** 567–70
- [70] Sepliarsky M and Cohen R E 2011 First-principles based atomistic modeling of phase stability in PMN–xPT *J. Phys. Condens. Matter* **23** 435902
- [71] Grinberg I and Rappe A M 2007 First principles calculations, crystal chemistry and properties of ferroelectric perovskites *Phase Transit.* **80** 351–68
- [72] Patel S, Chauhan A and Vaish R 2015 Mechanical confinement for tuning ferroelectric response in PMN-PT single crystal *J. Appl. Phys.* **117** 084102
- [73] Nadupalli S, Kreisel J and Granzow T 2019 Increasing bulk photovoltaic current by strain tuning *Sci. Adv.* **5** eaau9199
- [74] Yang -M-M, Kim D J and Alexe M 2018 Flexo-photovoltaic effect *Science* **360** 904–7
- [75] Liu X, Zhang F, Long P, Lu T, Zeng H, Liu Y, Withers R L, Li Y and Yi Z 2018 Anomalous photovoltaic effect in centrosymmetric ferroelastic BiVO$_4$ *Adv. Mater.* **30** 1801619
- [76] Sapper E, Dittmer R, Damjanovic D, Erdem E, Keeble D J, Jo W, Granzow T and Rödel J 2014 Aging in the relaxor and ferroelectric state of Fe-doped (1-x)(Bi$_{1/2}$Na$_{1/2}$)TiO$_3$-xBaTiO$_3$ piezoelectric ceramics *J. Appl. Phys.* **116** 104102
- [77] Hou W, Chowdhury S A, Dey A, Watson C, Peña T, Azizimanesh A, Askari H and Wu S M 2022 Nonvolatile ferroelastic strain from flexoelectric internal bias engineering *Phys. Rev. Appl.* **17** 024013
- [78] Takahashi S 1982 Effects of impurity doping in lead zirconate-titanate ceramics *Ferroelectrics* **41** 143–56
- [79] Huang Q *et al* 2022 Significant modulation of ferroelectric photovoltaic behavior by a giant macroscopic flexoelectric effect induced by strain-relaxed epitaxy *Adv. Electron. Mater.* **8** 2100612
- [80] Schlom D, Chen L-Q, Eom C-B, Rabe K, Streiffer S and Triscone J-M 2007 Strain tuning of ferroelectric thin films* *Annu. Rev. Mater. Res.* **37** 589–626